\documentclass{article}
\usepackage{amssymb}
\usepackage{amsfonts}
\usepackage{amsmath,indentfirst}
\usepackage[inner=2cm,outer=2cm,tmargin=1.5cm,bmargin=2cm,centering,vcentering]{geometry}
\usepackage{graphicx}
\usepackage{mathptmx}

\RequirePackage[colorlinks,citecolor=blue,urlcolor=blue,linkcolor=blue]{hyperref}
\newtheorem{theorem}{Theorem}

\input{tcilatex}
\begin{document}

\title{Generalized Self-similar Scalar-Tensor Theories }
\author{J.A. Belinch\'{o}n. \\
Departamento de F\'{\i}sica At\'{o}mica, Molecular y Nuclear, Ciencias F%
\'{\i}sicas. \\
Universidad Complutense de Madrid, E-28040 Madrid, Espa\~{n}a}
\maketitle

\begin{abstract}
We study through symmetry principles the form of the functions in the
generalized scalar-tensor theories under the self-similar hypothesis. The
results obtained are absolutely general and valid for all the Bianchi models
and the flat FRW one. We study the concrete example of the Kantowsky-Sach
model finding some new exact self-similar solutions.
\end{abstract}

\section{Introduction}

Current observations of the large scale cosmic microwave background suggest
to us that our physical universe is expanding in an accelerated way,
isotropic and homogeneous models with a positive cosmological constant. The
analysis of Cosmic Microwave Background (CMB) fluctuations could confirm
this picture. But other analyses reveal some inconsistencies. Analysis of
WMAP data sets shows us that the universe might have a preferred direction.
For this reason, it may be interesting to study Bianchi models since these
models may describe such anisotropies.

The observed location of the first acoustic peak of the temperature
fluctuations on the CMB corroborated by the data obtained in different
experiments \cite{bernadis}, indicates that the universe is dominated by an
unidentified \textquotedblleft dark energy\textquotedblright\ and suggests
that this unidentified dark energy has a negative pressure \cite{perlmutter}%
. This last characteristic of the dark energy points to the vacuum energy or
cosmological constant as a possible candidate for dark energy. Although, it
is a general belief that the current curvature of the universe is negligible
and mostly the universe is considered with a flat geometry, recent
observations support the possibility of a non-flat universe and detect a
small deviation from $k=0$ \cite{nonflat1}. For example, evidence from CMB
and also supernova measurements of the cubic correction to the luminosity
distance favour a positively curved universe \cite{nonflat2,nonflat3}.

In order to explain the current acceleration of the universe, within General
Relativity (GR), it is necessary to introduce a new type of energy with a
negative pressure. Between the different possible approaches is one which
consists of considering so-called dark energy (DE). There are several
candidates for DE, where the simplest one is the cosmological term $\Lambda $%
. However, this choice has several drawbacks such as coincidence and fine
tuning problems. For this reason other models have been proposed. Examples
of such models are quintessence \cite{wetter,ratra}, K-essence \cite{chiba},
or chameleonic fields in which a scalar field is coupled to matter \cite%
{chame}, \ etc. Hence it is natural and important to consider a variable $%
\Lambda -$term in more general frameworks where, furthermore, other
quantities, such as the Newton gravitational constant may be considered as
dynamical. One such class of theories are the scalar-tensor theories (STT)
of gravity. This class of models has received a renewed interest in recent
times, for two main reasons: First, the new inflationary scenario as the
extended inflation has a scalar field that solves several problems present
in the old theories. Secondly, string theories and other unified theories
contain a scalar field which plays a similar role to the scalar field of the
STT. The scalar-tensor theories started with the work of P. Jordan in 1950
\cite{jordan}. A prototype of such models was proposed by Brans and Dicke in
1961 \cite{BD}. Their aim for presenting this model was to modify Einstein's
theory in such a way as to incorporate the so called "Mach's principle".
These theories have been generalized by P.G. Bergmann \cite{Ber}, K.
Nordtverdt \cite{N} and R. T. Wagoner \cite{Wag}. For a recent review of
this class of theories we refer to \cite{Faraoni} and \cite{Fujii-Maeda}.

{\normalsize In this paper we want to consider a family of scalar-tensor
theories with a dynamical cosmological constant \cite{W} and with a
potential \cite{Faraoni}, that is equivalent to a time dependent
cosmological constant. }Recently several authors have considered the
cosmological consequences of a time varying cosmological constant. Most of
them introduce the time dependence in an ad hoc manner. In this work we
consider an equivalent problem in the well known general scalar-tensor
theory of gravity where the time dependence can occur in a natural way,
without any new assumption or modification of the theory and provide an
explanation for the acceleration of the universe expansion \cite{mathiaz}.

Many authors have studied these general scalar-tensor theories. They use the
observational data in order to obtain restrictions or constraint between the
functions that appear in the action to obtain an accelerated model \cite%
{Boisseau}-\cite{Bo2}. Our approach is different: We want to derive these
functions from symmetry principles as self-similarity. We shall carry out
our study under this assumption and state some general theorems that are
valid for all the Bianchi models and of course for the flat FRW one. It is
most appropriate for us to work in the Jordan frame (JF), in which the
physical quantities are those that are being measured in experiments, even
though the Einstein frame (EF) often provides a better mathematical insight.

The study of self-similar (SS) models is quite important since a large class
of orthogonal spatially homogeneous models are asymptotically self-similar
at the initial singularity and are approximated by exact perfect fluid or
vacuum self-similar power-law models. Exact self-similar power-law models
can also approximate general Bianchi models at intermediate stages of their
evolution. This last point is of particular importance in relating Bianchi
models to the real Universe. At the same time, self-similar solutions can
describe the behaviour of Bianchi models at late times i.e. as $t\rightarrow
\infty $ \cite{ColeyDS}.

This paper is organized as follows. In section two we start by considering a
particular formulation of the theory. In this case the cosmological constant
is introduced directly by the function $\Lambda \left( \phi \right) $ \cite%
{W}. We study this model through two different approaches. The first
consists of studying the effective stress-energy tensor under the matter
collineation approach. This method allows us to obtain relationships between
the physical quantities as well as to determine the exact form of the scalar
field $\phi .$ The second approach consists of studying the wave equation
under the Lie group method. By imposing a particular symmetry we are able to
determine the exact form for each of the functions that appear in this
equation. We summarize all the results by stating a very general theorem. In
section three we study a very general scalar-tensor theory. In this case the
cosmological constant is introduced by the potential. Following the same
exposed procedure as in the above section, we are able to determine the
exact form that all of the unknowns involved in this model must have. We
show, from the stated theorem, how different versions of this theory arise,
which are the standard Brans-Dicke theory, the induced gravity model \cite{z}
and a very particular solution where the effective gravitational function is
constant. In section four we study a chameleon Jordan-Brans-Dicke model. In
order to show how  all the obtained results work, in section five we study a
particular example which is the Kantowski-Sach model. We start this section
by showing that this metric admits a homothetic vector fields and then we
study several models. We put special emphasis on comparing the solutions
obtained in each case. In section six we end by summarizing all the results.
We have added an appendix where we study in detail one of the equations
obtained in section 2 in order to show with out any doubt that the
Brans-Dicke parameter $\omega \left( \phi \right) $ must be constant in this
framework of self-similar solutions.

\section{Cosmological models with dynamical $\Lambda $ in scalar-tensor
theories}

Following to Will (see \cite{W}) we start with the action for the most
general scalar-tensor theory of gravitation%
\begin{equation}
S=\frac{c^{3}}{16\pi G_{N}}\int d^{4}x\sqrt{-g}\left[ \phi R-\frac{\omega
\left( \phi \right) g^{ij}\phi _{,i}\phi _{,j}}{\phi }+2\phi \Lambda \left(
\phi \right) \right] +S_{NG},  \label{w1}
\end{equation}%
where $g=\det (g_{ij})$, $G_{\ast }$ is Newton's constant, $S_{NG}$ is the
action for the nongravitational matter. We use the signature ($-,+,+,+$).
The arbitrary functions $\omega \left( \phi \right) $ and $\Lambda \left(
\phi \right) $ distinguish the different scalar-tensor theories of
gravitation, $\Lambda \left( \phi \right) $ is a potential function and
plays the role of a cosmological constant, and $\omega \left( \phi \right) $
is the coupling function of the particular theory.

The explicit field equations are%
\begin{equation}
R_{ij}-\frac{1}{2}g_{ij}R=\frac{8\pi }{c^{4}\phi }T_{ij}+\Lambda \left( \phi
\right) g_{ij}+\frac{\omega }{\phi ^{2}}\left( \phi _{,i}\phi _{,j}-\frac{1}{%
2}g_{ij}\phi _{,l}\phi ^{,l}\right) +\frac{1}{\phi }\left( \phi
_{;ij}-g_{ij}\square \phi \right) ,  \label{w2}
\end{equation}

\begin{equation}
\left( 3+2\omega \left( \phi \right) \right) \square \phi =8\pi T-\frac{%
d\omega }{d\phi }\phi _{,l}\phi ^{,l}-2\phi \left( \phi \frac{d\Lambda }{%
d\phi }-\Lambda \left( \phi \right) \right) ,  \label{w4}
\end{equation}%
where $T=T_{i}^{i}$ is the trace of the stress-energy tensor. The
gravitational coupling $G_{\mathrm{eff}}(t)$ is given by

\begin{equation}
G_{\mathrm{eff}}(t)=\left( \frac{2\omega +4}{2\omega +3}\right) \frac{%
G_{\ast }}{\phi (t)}.  \label{G12}
\end{equation}

\subsection{Matter collineations}

We may calculate the relationship between the quantities (in a SS approach)
by calculating the matter collineations. Therefore we have to compute (we
use unit where $8\pi =c=1)$%
\begin{equation}
T_{ij}^{\mathrm{eff}}=\frac{1}{\phi }T_{ij}+\frac{\omega \left( \phi \right)
}{\phi ^{2}}\left( \phi _{,i}\phi _{,j}-\frac{1}{2}g_{ij}\phi _{,l}\phi
^{,l}\right) +\frac{1}{\phi }\left( \phi _{;ij}-g_{ij}\square \phi \right)
+\Lambda \left( \phi \right) g_{ij},
\end{equation}%
\begin{equation}
\mathcal{L}_{HO}\left( T_{ij}^{\mathrm{eff}}\right) =0,
\end{equation}%
where $HO$ stands for a homothetic vector field. For simplicity we have used
a flat FRW metric but we would like to emphasize that all the obtained
results are absolutely valid for all the Bianchi models, since we only look
for the behaviour of the physical quantities instead of restriction on the
scale factors. Therefore the homothetic vector field (HFV) is%
\begin{equation}
HO=\left( t+t_{0}\right) \partial _{t}+\left( 1-\left( t+t_{0}\right)
H\right) x\partial _{x}+\left( 1-\left( t+t_{0}\right) H\right) y\partial
_{y}+\left( 1-\left( t+t_{0}\right) H\right) z\partial _{z}.
\end{equation}%
Note the non-singular character of the HVF, nevertheless for simplicity in
the calculations we use the singular case.

\begin{enumerate}
\item $T_{1}=\phi ^{-1}T_{ij},$%
\begin{equation}
\mathcal{L}_{HO}\left( \frac{1}{\phi }T_{ij}\right) =0,\qquad
\Longleftrightarrow \qquad -t\rho \phi ^{\prime }+t\rho ^{\prime }\phi
+2\rho \phi =0,
\end{equation}%
obtaining%
\begin{equation}
\frac{\rho ^{\prime }}{\rho }-\frac{\phi ^{\prime }}{\phi }=-\frac{2}{t}%
\,\qquad \Longleftrightarrow \qquad \frac{\rho }{\phi }=t^{-2}.
\end{equation}

\item $T_{2}=\omega \left( \phi \right) \phi ^{-2}\left( \phi _{,i}\phi
_{,j}-\frac{1}{2}g_{ij}\phi _{,l}\phi ^{,l}\right) $%
\begin{equation}
\mathcal{L}_{HO}\left( \frac{\omega \left( \phi \right) }{\phi ^{2}}\left(
\phi _{,i}\phi _{,j}-\frac{1}{2}g_{ij}\phi _{,l}\phi ^{,l}\right) \right)
=0\qquad \Longleftrightarrow \qquad t\phi _{t}^{2}\left( \omega _{\phi }\phi
-2\omega \right) +2t\omega \phi _{tt}\phi +2\omega \phi _{t}\phi =0,
\end{equation}%
and therefore%
\begin{equation}
\phi _{tt}=-\frac{\phi _{t}^{2}}{\phi }\left( \frac{\omega _{\phi }\phi }{%
2\omega }-1\right) -\frac{\phi _{t}}{t},  \label{Eq phi}
\end{equation}%
or%
\begin{equation}
\frac{\omega _{\phi }}{\omega }\phi _{t}-2\frac{\phi _{t}}{\phi }+2\frac{%
\phi _{tt}}{\phi _{t}}=-\frac{2}{t}\qquad \frac{\omega ^{\prime }}{\omega }-2%
\frac{\phi _{t}}{\phi }+2\frac{\phi _{tt}}{\phi _{t}}=-\frac{2}{t}\,\qquad
\Longleftrightarrow \qquad \omega \left( \phi \right) \frac{\phi _{t}^{2}}{%
\phi ^{2}}=t^{-2}.
\end{equation}%
note that $\omega ^{\prime }=\omega _{\phi }\phi _{t}.$

We find the next solution for Eq. (\ref{Eq phi})%
\begin{equation}
\int^{\phi }\frac{\sqrt{\omega \left( \varphi \right) }}{\varphi }d\varphi
-C_{1}\ln t+C_{2}=0.  \label{eva1}
\end{equation}

For example

\begin{itemize}
\item $\omega \left( \varphi \right) =cont$%
\begin{equation}
\int^{\phi }\frac{k}{\varphi }d\varphi =k\ln \phi ,\qquad \Longrightarrow
\qquad \phi =\phi _{0}t^{n},
\end{equation}%
this is the unique solution mathematically possible (compatible with the SS
hypothesis, see the appendix for an explanation).

\item $\omega \left( \varphi \right) =\varphi ^{a}$%
\begin{equation}
\int^{\phi }\frac{\varphi ^{a/2}}{\varphi }d\varphi =\frac{2}{a}\phi
^{a/2}\qquad \Longrightarrow \qquad \frac{2}{a}\phi ^{a/2}=C_{1}\ln t\qquad
\phi =C_{1}\left( \ln t\right) ^{2/a}.
\end{equation}
\end{itemize}

If\textbf{\ }$\omega =const.$\textbf{\ }then%
\begin{equation}
t\phi _{t}^{2}-t\phi _{tt}\phi -\phi _{t}\phi =0,
\end{equation}%
and therefore we get the following ODE%
\begin{equation}
\phi _{tt}=\frac{\phi _{t}^{2}}{\phi }-\frac{\phi _{t}}{t}\qquad
\Longleftrightarrow \qquad \phi =\exp \left( -C_{3}-C_{2}\ln t\right) =\phi
_{0}t^{C_{2}}.  \label{phi_sol}
\end{equation}

\item $T_{3}=\phi ^{-1}\left( \phi _{;ij}-g_{ij}\square \phi \right) $%
\begin{equation}
\mathcal{L}_{HO}\left( \frac{1}{\phi }\left( \phi _{;ij}-g_{ij}\square \phi
\right) \right) =0,
\end{equation}%
i.e.%
\begin{eqnarray}
t\left( \phi _{tt}f^{\prime }+\phi _{t}f^{\prime \prime }-\frac{\phi _{t}^{2}%
}{\phi }f^{\prime }-\phi _{t}\frac{f^{\prime 2}}{f}\right) +2\phi
_{t}f^{\prime } &=&0, \\
t\left[ \phi ^{\prime \prime \prime }+\left( 2H-\frac{\phi ^{\prime }}{\phi }%
\right) \phi ^{\prime \prime }-2H\frac{\phi ^{\prime ^{2}}}{\phi }%
+2H^{\prime }\phi ^{\prime }\right] +2\left( \phi ^{\prime \prime }-2\phi
^{\prime }H\right) &=&0,
\end{eqnarray}%
note that $H=ht^{-1},$ $h\in \mathbb{R}^{+}.$ These equation are different
for each Bianchi model and we only obtain restriction on the scale factors.

\item $T_{4}=\Lambda \left( \phi \right) g_{ij},$%
\begin{equation}
\mathcal{L}_{HO}\left( \Lambda \left( \phi \right) g_{ij}\right) =0
\end{equation}%
i.e.%
\begin{equation}
t\Lambda _{\phi }\phi ^{\prime }+2\Lambda =0\,\qquad \Longleftrightarrow
\qquad \frac{\Lambda _{\phi }}{\Lambda }\phi ^{\prime }=-\frac{2}{t}\qquad
\Longleftrightarrow \qquad \Lambda =\Lambda _{0}t^{-2},
\end{equation}%
where $\Lambda ^{\prime }=\Lambda _{\phi }\phi ^{\prime }.$
\end{enumerate}

\subsection{Lie groups}

We are going to study the Eq. (\ref{w4}) through the LG method, i.e. we
study the kind of functions $\Lambda \left( \phi \right) $ and $\omega
\left( \phi \right) $ such that this equation is integrable. We start by
rewriting it in an appropriate way%
\begin{equation}
\left( 3+2\omega \left( \phi \right) \right) \left( \phi ^{\prime \prime
}+ht^{-1}\phi ^{\prime }\right) =Ct^{-\alpha }+B\left( \Lambda -\phi \Lambda
_{\phi }\right) \phi -\phi ^{\prime 2}\omega _{\phi },
\end{equation}%
where $h=\mathrm{const.},$ $h\in \mathbb{R}^{+},$ $B=2,$ and $C=8\pi \left(
1-3\gamma \right) \rho _{0}.$ Note that we are taking into account the
conservation equation $\func{div}T=0,$ i.e. $\rho =\rho _{0}t^{-\alpha },$
where $\alpha =(1+\gamma )h,$ and $H=ht^{-1}.$

We need to solve the following system of PDE%
\begin{eqnarray}
\omega _{\phi }\xi _{\phi }-W\xi _{\phi \phi } &=&0,  \label{Je01} \\
2ht^{-1}W\xi _{\phi }+W\eta _{\phi \phi }-2W\xi _{\phi t}-\left(
2W^{-1}\omega _{\phi }^{2}-\omega _{\phi \phi }\right) \eta +\omega _{\phi
}\eta _{\phi } &=&0,  \label{Je02}
\end{eqnarray}%
\begin{equation*}
-3\left( B\phi \left( \Lambda -\phi \Lambda _{\phi }\right) +Ct^{-\alpha
}\right) \xi _{\phi }+ht^{-2}W\left( t\xi _{t}-\xi \right) +2W\eta _{t\phi }-
\end{equation*}%
\begin{equation}
-W\xi _{tt}+2ht^{-1}\omega _{\phi }\left( 1-\left( 3+2\omega \right)
W^{-1}\right) \eta +2\omega _{\phi }\eta _{t}=0,  \label{he1}
\end{equation}%
\begin{equation*}
\left[ B\left( \phi ^{2}\Lambda _{\phi \phi }+\phi \Lambda _{\phi }-\Lambda
\right) +2\omega _{\phi }W^{-1}\left( B\phi \left( \Lambda -\phi \Lambda
_{\phi }\right) +Ct^{-\alpha }\right) \right] \eta -2\left[ B\phi \left(
\Lambda -\phi \Lambda _{\phi }\right) -Ct^{-\alpha }\right] \xi _{t}+
\end{equation*}%
\begin{equation}
+\alpha Ct^{-\alpha -1}\xi +\left[ B\phi \left( \Lambda -\phi \Lambda _{\phi
}\right) +Ct^{-\alpha }\right] \eta _{\phi }+ht^{-1}W\eta _{t}+W\eta _{tt}=0,
\label{he2}
\end{equation}%
where $W=\left( 3+2\omega \left( \phi \right) \right) .$ notice that $\left(
2\omega +3\right) W^{-1}=1,$ so Eq. (\ref{he1}) yields%
\begin{equation}
-3\left( B\phi \left( \Lambda -\phi \Lambda _{\phi }\right) +Ct^{-\alpha
}\right) \xi _{\phi }+ht^{-2}W\left( t\xi _{t}-\xi \right) +2W\eta _{t\phi
}-W\xi _{tt}+2\omega _{\phi }\eta _{t}=0.  \label{he3}
\end{equation}

The symmetry $\xi =t,\eta =n\phi $, brings us to obtain the following
restriction on $\Lambda \left( \phi \right) .$ From Eq. (\ref{he2}) we get%
\begin{equation*}
\left[ B\left( \phi ^{2}\Lambda _{\phi \phi }+\phi \Lambda _{\phi }-\Lambda
\right) +2\omega _{\phi }W^{-1}\left( B\phi \left( \Lambda -\phi \Lambda
_{\phi }\right) +Ct^{-\alpha }\right) \right] n\phi -2\left[ B\phi \left(
\Lambda -\phi \Lambda _{\phi }\right) -Ct^{-\alpha }\right] +
\end{equation*}%
\begin{equation}
+\alpha Ct^{-\alpha }+\left[ B\phi \left( \Lambda -\phi \Lambda _{\phi
}\right) +Ct^{-\alpha }\right] n=0,
\end{equation}%
so%
\begin{equation}
Ct^{-\alpha }\left( \alpha +n-2+2n\omega _{\phi }\phi W^{-1}\right) =0,
\label{zoe1}
\end{equation}%
and
\begin{equation}
n\left( \phi ^{2}\Lambda _{\phi \phi }+\phi \Lambda _{\phi }-\Lambda \right)
+\left( \Lambda -\phi \Lambda _{\phi }\right) \left( -\alpha \right) =0,
\end{equation}%
where we have taken into account Eq. (\ref{zoe1}) therefore we have obtained
the following ODE for $\Lambda \left( \phi \right) $%
\begin{equation}
\Lambda _{\phi \phi }=\left( \frac{n+\alpha }{n}\right) \left( -\frac{%
\Lambda _{\phi }}{\phi }+\frac{\Lambda }{\phi ^{2}}\right) ,
\end{equation}%
and whose general solution is%
\begin{equation}
\Lambda \left( \phi \right) =\Lambda _{0}\phi ^{\frac{-1}{n}\left( n+\alpha
\right) }+\frac{C_{3}n}{2n+\alpha }\phi ,
\end{equation}%
we choose
\begin{equation}
\Lambda \left( \phi \right) =\Lambda _{0}\phi ^{\frac{-1}{n}\left( n+\alpha
\right) }=\Lambda _{0}t^{-\left( n+\alpha \right) },  \label{CC}
\end{equation}%
since from our result from the MC approach we already know that, $\Lambda
\left( t\right) =t^{-2},$ so this means that $n+\alpha =2.$

In the same way we may calculate the restriction on $\omega \left( \phi
\right) ,$%
\begin{equation}
\omega _{\phi }\phi W^{-1}=\frac{2-\alpha -n}{2n},\qquad \qquad \omega
^{\prime }=\left( \frac{2-\alpha -n}{2n}\right) \frac{\left( 3+2\omega
\right) }{\phi },
\end{equation}%
whose solutions are
\begin{equation}
\omega \left( \phi \right) =\phi ^{\frac{-1}{n}\left( n+\alpha -2\right)
}e^{-2nC_{4}}-\frac{3}{2}=\omega _{n}t^{-\left( n+\alpha -2\right) }-\frac{3%
}{2},\qquad \omega _{n}=\mathrm{const},
\end{equation}%
therefore we obtain
\begin{equation}
\omega \left( \phi \right) =const.,\qquad n+\alpha =2.
\end{equation}

In an alternative way, from Eq. (\ref{Je02}) we get%
\begin{equation}
\omega _{\phi \phi }=\left( \frac{2-\alpha -2n}{n}\right) \frac{\omega
_{\phi }}{\phi },\qquad \Longrightarrow \qquad \omega \left( \phi \right)
=C_{8}+C_{9}\phi ^{\frac{1}{n}\left( 2-n-\alpha \right) }=\omega
_{0}t^{2-n-\alpha },
\end{equation}%
and therefore%
\begin{equation}
\omega \left( \phi \right) =\omega _{0}=\mathrm{const},
\end{equation}%
since $2=n+\alpha .$ In the appendix we shall give and alternative and
detailed proof of this result.

\begin{theorem}
The scaling and in particular the self-similar solution admitted for the FE (%
\ref{w2}-\ref{w4}) have the following form%
\begin{equation*}
\phi =\phi _{0}\left( t+t_{0}\right) ^{n},\qquad \Lambda \left( \phi \right)
=\Lambda _{0}\phi ^{\frac{-1}{n}\left( n+\alpha \right) }=\Lambda _{0}\left(
t+t_{0}\right) ^{-\left( n+\alpha \right) },
\end{equation*}%
$\qquad $\ with $n+\alpha =2,$ therefore $\Lambda \left( t\right) =\Lambda
_{0}\left( t+t_{0}\right) ^{-2}.$ The Brans-Dicke parameter is constant%
\begin{equation*}
\omega \left( \phi \right) =\mathrm{const}.
\end{equation*}%
and $\rho =\rho _{0}\left( t+t_{0}\right) ^{-\alpha },$ $\alpha =\left(
1+\gamma \right) h.$
\end{theorem}

\section{The General case}

We start by defining the action \cite{F}

\begin{equation}
S={\frac{1}{8\pi }}\int d^{4}x\sqrt{-g}\left\{ \frac{1}{2}\left[ F\left(
\phi \right) R-Z\left( \phi \right) \phi _{,\alpha }\phi _{,}^{\alpha
}-2U(\phi )\right] +\mathcal{L}_{M}\right\} ,  \label{AG}
\end{equation}%
and therefore the FE read%
\begin{equation}
FG_{\mu \nu }=8\pi T_{\mu \nu }^{Matter}+Z\left[ \phi _{;\mu }\phi _{;\nu }-%
\frac{1}{2}g_{\mu \nu }\phi _{;\alpha }\phi _{;}^{\alpha }\right] +\left[
F_{;\mu }F_{;\nu }-g_{\mu \nu }\square F\right] -U(\phi )g_{\mu \nu }\,,
\label{Enma1}
\end{equation}%
and%
\begin{equation}
2Z\square \phi =F_{\phi }R-Z_{\phi }\phi ^{\prime 2}-2U_{\phi }\,,
\label{Enma2}
\end{equation}%
where $R$ is the scalar curvature.

The effective gravitational constant $G_{\mathrm{eff}}$ between two test
masses measured in laboratory Cavendish-type experiments is given by
\begin{equation}
G_{\mathrm{eff}}=\frac{G_{\ast }}{F}\left[ \frac{2Z(\phi )F+4F_{\phi }^{2}}{%
2Z(\phi )F+3F_{\phi }^{2}}\right] \,.  \label{Geff01}
\end{equation}

\subsection{Matter collienations}

We may define%
\begin{equation}
T_{ij}^{\mathrm{eff}}=\frac{1}{F\left( \phi \right) }T_{ij}+\frac{Z\left(
\phi \right) }{F\left( \phi \right) }\left( \phi _{,i}\phi _{,j}-\frac{1}{2}%
g_{ij}\phi _{,l}\phi ^{,l}\right) +\frac{1}{F\left( \phi \right) }\left(
F_{;i}F_{;j}-g_{ij}\square F\right) -\frac{U(\phi )}{F\left( \phi \right) }%
g_{ij}\,,
\end{equation}%
and therefore, as above:

\begin{enumerate}
\item $T_{1}=F\left( \phi \right) ^{-1}T_{ij}$%
\begin{equation}
\mathcal{L}_{HO}\left( F\left( \phi \right) ^{-1}T_{ij}\right) =0\qquad
\Longleftrightarrow \qquad t\rho F_{\phi }\phi ^{\prime }-t\rho ^{\prime
}F-2\rho F=0,
\end{equation}%
so%
\begin{equation}
\frac{F_{\phi }}{F}\phi ^{\prime }-\frac{\rho ^{\prime }}{\rho }=\frac{-2}{t}%
\,\qquad \Longleftrightarrow \qquad \frac{\rho }{F}=t^{-2}.
\end{equation}

\item $T_{2}=\frac{Z\left( \phi \right) }{F\left( \phi \right) }\left( \phi
_{,i}\phi _{,j}-\frac{1}{2}g_{ij}\phi _{,l}\phi ^{,l}\right) $%
\begin{equation}
\mathcal{L}_{HO}\left( \frac{Z\left( \phi \right) }{F\left( \phi \right) }%
\left( \phi _{,i}\phi _{,j}-\frac{1}{2}g_{ij}\phi _{,l}\phi ^{,l}\right)
\right) =0\qquad \Longleftrightarrow \qquad t\rho FZ_{\phi }\phi ^{\prime
2}-tZF_{\phi }\phi ^{\prime 2}+2tZF\phi ^{\prime \prime }+2ZF\phi ^{\prime
}=0,
\end{equation}%
thus%
\begin{equation}
\frac{Z_{\phi }}{Z}\phi ^{\prime }-\frac{F_{\phi }}{F}\phi ^{\prime }+2\frac{%
\phi ^{\prime \prime }}{\phi ^{\prime }}=\frac{-2}{t}\,\qquad
\Longleftrightarrow \qquad \frac{Z}{F}\phi ^{\prime 2}=t^{-2}.
\end{equation}

\item $T_{3}=\frac{1}{F\left( \phi \right) }\left( F_{;\mu }F_{;\nu }-g_{\mu
\nu }\square F\right) $ depends on the metric so we only obtain restriction
on the scale factors.

\item $T_{4}=\frac{U(\phi )}{F\left( \phi \right) }g_{\mu \nu }$%
\begin{equation}
\mathcal{L}_{HO}\left( \frac{U(\phi )}{F\left( \phi \right) }g_{\mu \nu
}\right) =0\qquad \Longleftrightarrow \qquad t\phi ^{\prime }U_{\phi
}F-tUF_{\phi }\phi ^{\prime }+2\phi U=0,
\end{equation}%
and therefore%
\begin{equation}
\frac{U_{\phi }}{U}\phi ^{\prime }-\frac{F_{\phi }}{F}\phi ^{\prime }=\frac{%
-2}{t}\,\qquad \Longleftrightarrow \qquad \frac{U}{F\left( \phi \right) }%
=t^{-2}.
\end{equation}
\end{enumerate}

\subsection{Lie groups}

For this model we have to solve the following equation%
\begin{equation}
2Z\square \phi =F_{\phi }R-Z_{\phi }\phi ^{\prime 2}-2U_{\phi },
\end{equation}%
that we may rewrite in the following form%
\begin{equation}
2Z\left( \phi ^{\prime \prime }+ht^{-1}\phi ^{\prime }\right)
=Ct^{-2}F_{\phi }-Z_{\phi }\phi ^{\prime 2}-2U_{\phi },
\end{equation}%
where we have assumed $\phi =\phi (t)$, and the derivatives respect $t$ are
denoted by a comma. Note that $R\thickapprox Ct^{-2},$ with $C\in \mathbb{R}%
. $

The standard procedure brings us to outline the following system of PDE:%
\begin{eqnarray}
Z_{\phi }\xi _{\phi }-2Z\xi _{\phi \phi } &=&0,  \label{Ash01} \\
\left( ZZ_{\phi \phi }-Z_{\phi }^{2}\right) \eta +ZZ_{\phi }\eta _{\phi
}+4Z^{2}ht^{-1}\xi _{\phi }+2Z^{2}\eta _{\phi \phi }-4Z^{2}\xi _{t\phi }
&=&0,  \label{Ash02} \\
3\left( 2V_{\phi }-Ct^{-2}F_{\phi }\right) \xi _{\phi }+2ht^{-2}Z\left( t\xi
_{t}-\xi \right) +4Z\eta _{t\phi }-2Z\xi _{tt}+2Z_{\phi }\eta _{t} &=&0,
\label{Ash03}
\end{eqnarray}%
\begin{equation*}
\left[ Ct^{-2}\left( Z_{\phi }F_{\phi }-ZF_{\phi \phi }\right) +2\left(
ZU_{\phi \phi }-Z_{\phi }U_{\phi }\right) \right] \eta +Z\left(
Ct^{-2}F_{\phi }-2U_{\phi }\right) \eta _{\phi }+
\end{equation*}%
\begin{equation}
2Z\left( 2U_{\phi }-Ct^{-2}F_{\phi }\right) \xi _{t}+2Ct^{-3}ZF_{\phi }\xi
+2Z^{2}\left( ht^{-1}\eta _{t}+\eta _{tt}\right) =0.  \label{Ash04}
\end{equation}

The symmetry $\xi =t,\eta =n\phi $, brings us to obtain the following
restrictions. From Eq. (\ref{Ash02})
\begin{equation}
Z_{\phi \phi }=\frac{Z_{\phi }^{2}}{Z}-Z_{\phi }\frac{\eta _{\phi }}{\eta }%
,\qquad \Longrightarrow \qquad Z_{\phi \phi }=\frac{Z_{\phi }^{2}}{Z}-\frac{%
Z_{\phi }}{\phi },\qquad \Longrightarrow \qquad Z(\phi )=Z_{0}\phi ^{-m},
\end{equation}%
where $m\in \mathbb{R}.$ From Eq. (\ref{Ash04}) we get%
\begin{eqnarray}
2\left( ZU_{\phi \phi }-Z_{\phi }U_{\phi }\right) \eta -2ZU_{\phi }\eta
_{\phi }+4ZU_{\phi }\xi _{t} &=&0, \\
Ct^{-2}\left[ \left( Z_{\phi }F_{\phi }-ZF_{\phi \phi }\right) \eta
+ZF_{\phi }\eta _{\phi }-2ZF_{\phi }\xi _{t}+2t^{-1}ZF_{\phi }\xi \right]
&=&0,
\end{eqnarray}%
and therefore
\begin{equation}
U_{\phi \phi }=U_{\phi }\frac{Z_{\phi }}{Z}+\left( \frac{n-2}{n}\right)
\frac{U_{\phi }}{\phi },\qquad U_{\phi \phi }=\left( \frac{n-2}{n}-m\right)
\frac{U_{\phi }}{\phi },
\end{equation}%
so%
\begin{equation}
U\left( \phi \right) =C_{2}+U_{0}\phi ^{-\frac{1}{n}\left( mn-2n+2\right) },
\end{equation}%
while%
\begin{equation}
F_{\phi \phi }=F_{\phi }\frac{Z_{\phi }}{Z}+\frac{F_{\phi }}{\phi },\qquad
F_{\phi \phi }=\left( 1-m\right) \frac{F_{\phi }}{\phi },
\end{equation}%
obtaining%
\begin{equation}
F\left( \phi \right) =C_{1}+F_{0}\phi ^{2-m}.
\end{equation}

\begin{theorem}
The scaling and in particular the self-similar solution admitted for the FE (%
\ref{Enma1}-\ref{Enma2}) have the following form%
\begin{equation}
\phi =\phi _{0}\left( t+t_{0}\right) ^{n},\qquad Z(\phi )=Z_{0}\phi
^{-m}=Z_{0}\left( t+t_{0}\right) ^{-nm},
\end{equation}%
with%
\begin{equation}
\rho =\rho _{0}\left( t+t_{0}\right) ^{-\alpha },\qquad \alpha =\left(
1+\gamma \right) h,
\end{equation}%
while%
\begin{equation}
F\left( \phi \right) =C_{1}+F_{0}\phi ^{2-m},\,\qquad U\left( \phi \right)
=C_{2}+U_{0}\phi ^{-\frac{1}{n}\left( mn-2n+2\right) }
\end{equation}
\end{theorem}

This theorem state that we have a set of theories which admit scaling and in
particular self-similar solutions if the involved functions take this
particular form. The action reads

\begin{equation}
S={\frac{1}{8\pi }}\int d^{4}x\sqrt{-g}\left\{ \frac{1}{2}\left[ \phi
^{2-m}R-\phi ^{-m}\phi _{,\alpha }\phi _{,}^{\alpha }-2U(\phi )\right] +%
\mathcal{L}_{M}\right\} .
\end{equation}%
So setting different values for the constant $m$ we obtain different
theories. For example if $m=0$ (induced gravity case) then the action (\ref%
{AG}) yields%
\begin{equation*}
S={\frac{1}{8\pi }}\int d^{4}x\sqrt{-g}\left\{ \frac{1}{2}\left[ \phi
^{2}R-\phi _{,\alpha }\phi _{,}^{\alpha }-2U(\phi )\right] +\mathcal{L}%
_{M}\right\} ,
\end{equation*}%
\begin{equation*}
\phi =\phi _{0}\left( t+t_{0}\right) ^{n},\qquad Z(\phi )=Z_{0},\qquad
F\left( \phi \right) =F_{0}\phi ^{2},\qquad U\left( \phi \right) =U_{0}\phi
^{\frac{1}{n}\left( 2n-2\right) },\qquad G_{\mathrm{eff}}\thickapprox \phi
^{-2},
\end{equation*}%
and if $m=1$ (usual JBD theory with a potential) then the action (\ref{AG})
yields%
\begin{equation*}
S={\frac{1}{8\pi }}\int d^{4}x\sqrt{-g}\left\{ \frac{1}{2}\left[ \phi R-%
\frac{\omega }{\phi }\phi _{,\alpha }\phi _{,}^{\alpha }-2U(\phi )\right] +%
\mathcal{L}_{M}\right\} .
\end{equation*}%
\begin{equation*}
\phi =\phi _{0}\left( t+t_{0}\right) ^{n},\qquad Z(\phi )=Z_{0}\phi
^{-1},\qquad F\left( \phi \right) =F_{0}\phi ,\qquad U\left( \phi \right)
=U_{0}\phi ^{\frac{1}{n}\left( n-2\right) },\qquad G_{\mathrm{eff}%
}\thickapprox \phi ^{-1},
\end{equation*}%
i.e.%
\begin{equation*}
Z(\phi )=\frac{\omega \left( \phi \right) }{\phi },\qquad \omega \left( \phi
\right) =\mathrm{const}.
\end{equation*}%
and to end if $m=2,$ then the action (\ref{AG}) yields%
\begin{equation*}
S={\frac{1}{8\pi }}\int d^{4}x\sqrt{-g}\left\{ \frac{1}{2}\left[ R-\frac{%
\omega }{\phi ^{2}}\phi _{,\alpha }\phi _{,}^{\alpha }-2U(\phi )\right] +%
\mathcal{L}_{M}\right\} .
\end{equation*}%
we get:
\begin{equation*}
\phi =\phi _{0}\left( t+t_{0}\right) ^{n},\qquad Z(\phi )=Z_{0}\phi
^{-2},\qquad F\left( \phi \right) =F_{0},\qquad U\left( \phi \right)
=U_{0}\phi ^{\frac{-2}{n}},\qquad G_{\mathrm{eff}}\thickapprox \mathrm{const}%
..
\end{equation*}%
This particular case is very similar to the scalar field cosmological model
with $G_{\mathrm{eff}}\thickapprox \mathrm{const}.$ Notice that this model
is quite different from the Barker's theory \cite{Barker}.

\section{Chameleon cosmology}

We begin with the BD chameleon theory in which the scalar field is coupled
non-minimally to the matter field via the action {\
\begin{equation}
S=\int {\ d^{4}x\sqrt{-g}\left( \phi {R}-\frac{\omega }{\phi }g^{\mu \nu
}\partial _{\mu }\phi \partial _{\nu }\phi -U(\phi )+J(\phi )L_{m}\right) },
\label{act1}
\end{equation}%
} where ${R}$ is the Ricci scalar curvature, $\phi $ is the BD scalar field {%
\ with a potential $U(\phi )$. The chameleon field $\phi $ is} non-minimally
coupled to gravity, $\omega $ is the dimensionless BD parameter. The last
term in the action indicates the interaction between the matter Lagrangian $%
L_{m}$ and some arbitrary function $J(\phi )$ of the BD scalar field. In the
limiting case $J(\phi )=1$, we obtain the standard BD theory.

{The gravitational field equations derived from the action (\ref{act1}) with
respect to the metric is
\begin{equation}
R_{\mu \nu }-\frac{1}{2}g_{\mu \nu }R=\frac{J(\phi )}{\phi }T_{\mu \nu }+%
\frac{\omega }{\phi ^{2}}\left( \phi _{\mu }\phi _{\nu }-\frac{1}{2}g_{\mu
\nu }\phi ^{\alpha }\phi _{\alpha }\right) +\frac{1}{\phi }[\phi _{\mu ;\nu
}-g_{\mu \nu }\Box \phi ]-g_{\mu \nu }\frac{U(\phi )}{2\phi }.  \label{olga}
\end{equation}%
}

{The Klein-Gordon equation (or the wave equation) for the scalar field is
\begin{equation}
\left( 2\omega +3\right) \Box \phi =T\left( J-\frac{1}{2}\phi J_{,\phi
}\right) +(\phi U_{,\phi }-2U).  \label{phi1}
\end{equation}%
}

{Similarly the energy conservation for the cosmic fluid is
\begin{equation}
\dot{\rho}+\theta (\rho +p)=0,\qquad \theta =u_{;i}^{i}.  \label{FE4}
\end{equation}%
We shall use the equation of state (EoS) for the fluid $p=\gamma \rho $,
thus (\ref{FE4}) yields $\rho ={\rho }_{0}\left( t+t_{0}\right) ^{-\alpha }.$%
}

\subsection{Matter collineations}

By defining%
\begin{equation}
T_{ij}^{\mathrm{eff}}=\frac{J(\phi )}{\phi }T_{\mu \nu }+\frac{\omega }{\phi
^{2}}\left( \phi _{\mu }\phi _{\nu }-\frac{1}{2}g_{\mu \nu }\phi ^{\alpha
}\phi _{\alpha }\right) +\frac{1}{\phi }[\phi _{\mu ;\nu }-g_{\mu \nu }\Box
\phi ]-g_{\mu \nu }\frac{U(\phi )}{2\phi },
\end{equation}%
as it is observed it is only necessary to calculate the first component, the
rest of them have been already calculated in the above sections.

If $T_{1}=F(\phi )\phi ^{-1}T_{ij},$ then%
\begin{equation}
\mathcal{L}_{HO}\left( \frac{J(\phi )}{\phi }T_{ij}\right) =0,
\end{equation}%
yields%
\begin{equation}
tJ_{\phi }\phi ^{\prime }\rho \phi -t\rho J\phi ^{\prime }+t\rho ^{\prime
}J\phi +2\rho J\phi =0,
\end{equation}%
algebra brings us to get:%
\begin{equation}
\frac{J_{\phi }\phi ^{\prime }}{J}+\frac{\rho ^{\prime }}{\rho }-\frac{\phi
^{\prime }}{\phi }=-\frac{2}{t}\,\qquad \Longleftrightarrow \qquad \frac{J}{%
\phi }\rho =t^{-2}.
\end{equation}%
where$J^{\prime }=J_{\phi }\phi ^{\prime },$ but we do not obtain more
information about the behaviour of some of the functions $\phi $ or $J\left(
\phi \right) .$

\subsection{Lie Groups}

{We need to study the following ODE{\ }}%
\begin{equation}
\phi ^{\prime \prime }+ht^{-1}\phi ^{\prime }=C\left( J-\frac{1}{2}\phi
J_{\phi }\right) t^{-\alpha }+K\left( 2V-\phi U_{\phi }\right) ,
\end{equation}%
where, $C=\frac{8\pi \left( 1-3\gamma \right) }{3+2\omega },$ $K=\frac{1}{%
3+2\omega },$ $h=\mathrm{const.},$

{Therefore the standard procedure brings us to outline the following system
of PDE}%
\begin{eqnarray}
\xi _{\phi \phi } &=&0, \\
ht^{-1}\xi _{\phi }+\eta _{\phi \phi }-2\xi _{\phi t} &=&0, \\
ht^{-2}\left( t\xi _{t}-\xi \right) +2\eta _{t\phi }-\xi _{tt}-3\left[
Ct^{-\alpha }\left( J-\frac{1}{2}\phi J_{\phi }\right) +2U-\phi U_{\phi }%
\right] \xi _{\phi } &=&0,
\end{eqnarray}%
\begin{equation*}
\eta _{tt}+ht^{-1}\eta _{t}+2\left( \phi U_{\phi }-U-Ct^{-\alpha }\left( J-%
\frac{1}{2}\phi J_{\phi }\right) \right) \xi _{t}+\alpha Ct^{-\alpha
-1}\left( J-\frac{1}{2}\phi J_{\phi }\right) \xi
\end{equation*}%
\begin{equation}
+\left( 2V-\phi V_{\phi }+Ct^{-\alpha }\left( J-\frac{1}{2}\phi J_{\phi
}\right) \right) \eta _{\phi }-\left( \frac{Ct^{-\alpha }}{2}\left( J_{\phi
}-\phi J_{\phi \phi }\right) +U_{\phi }-\phi U_{\phi \phi }\right) \eta =0,
\end{equation}

As we already know, from the MC approach, it must be verified the
relationship%
\begin{equation}
\frac{J\left( \phi \right) }{\phi }\rho =t^{-2},\qquad \Longrightarrow
\qquad \frac{J\left( \phi \right) }{\phi }=t^{-2+\alpha }.
\end{equation}%
algebra brings us to get%
\begin{eqnarray}
Ct^{-\alpha }\left[ \left( J-\frac{1}{2}\phi J_{\phi }\right) \left( \alpha
t^{-1}\xi -2\xi _{t}+\eta _{\phi }\right) -\frac{1}{2}\left( J_{\phi }-\phi
J_{\phi \phi }\right) \eta \right] &=&0,  \label{ang1} \\
\left( U_{\phi }-\phi U_{\phi \phi }\right) \eta +\left( 2U-\phi U_{\phi
}\right) \left( -2\xi _{t}+\eta _{\phi }\right) &=&0.  \label{ang2}
\end{eqnarray}

For example the symmetry $\xi =t,\eta =n\phi $, brings us to obtain the
following restriction on $U\left( \phi \right) .$ From Eq. (\ref{ang2}) we
obtain the ODE%
\begin{equation}
U_{\phi \phi }=2\left( 1-\frac{1}{n}\right) \frac{U_{\phi }}{\phi }+2\left(
\frac{2}{n}-1\right) \frac{U}{\phi ^{2}},
\end{equation}%
whose solution has been obtained in section 3, i.e. \ $U\left( \phi \right)
=U_{0}\phi ^{\frac{1}{n}\left( n-2\right) }.$ From Eq. (\ref{ang1}) we get
the next ODE for $J:$%
\begin{equation}
J_{\phi \phi }=\left( \frac{\alpha +2n-2}{n}\right) \frac{J_{\phi }}{\phi }%
-2\left( \frac{\alpha +n-2}{n}\right) \frac{J}{\phi ^{2}},
\end{equation}%
finding that the most general solution is%
\begin{equation}
J\left( \phi \right) =J_{0}\phi ^{\frac{1}{n}\left( n+\alpha -2\right)
}+C_{2}\phi ^{2}.
\end{equation}

\begin{theorem}
The scaling and in particular the self-similar solution admitted for the FE (%
\ref{Enma1}-\ref{Enma2}) have the following form%
\begin{equation}
\phi =\phi _{0}\left( t+t_{0}\right) ^{n}\qquad \Longrightarrow \qquad
U\left( \phi \right) =U_{0}\phi ^{\frac{1}{n}\left( n-2\right) }=U_{0}\left(
t+t_{0}\right) ^{n-2},
\end{equation}%
with $\rho =\rho _{0}\left( t+t_{0}\right) ^{-\alpha },$ $\alpha =\left(
1+\gamma \right) h,$ while%
\begin{equation}
J\left( \phi \right) =J_{0}\phi ^{\frac{1}{n}\left( n+\alpha -2\right)
}=J_{0}\left( t+t_{0}\right) ^{n+\alpha -2}.
\end{equation}
\end{theorem}

\section{The Kantowski-Sach model}

We start by considering the Killing vector fields (KVF)
\begin{equation}
Z_{1}=\partial _{y},\quad Z_{2}=\cot z\cos y\partial _{y}+\sin y\partial
_{z},\quad Z_{3}=-\cot z\sin y\partial _{y}+\cos y\partial _{z},\quad
Z_{4}=\partial _{x},
\end{equation}%
such that
\begin{equation}
\left[ Z_{1},Z_{2}\right] =Z_{3},\quad \left[ Z_{2},Z_{3}\right]
=Z_{1},\quad \left[ Z_{3},Z_{1}\right] =Z_{2},\quad \left[ Z_{4},Z_{i}\right]
=0,
\end{equation}%
in such a way that the metric takes the following form
\begin{equation}
ds^{2}=-dt^{2}+a^{2}(t)dx^{2}+b^{2}(t)\left( \sin ^{2}zdy^{2}+dz^{2}\right) ,
\label{ksmetric}
\end{equation}

We find that the metric (\ref{ksmetric}) admits the following HVF
\begin{equation}
H=\left( t+t_{0}\right) \partial_{t}+\left( 1-\left( t+t_{0}\right) \frac{%
a^{\prime}}{a}\right) x\partial_{x}+\left( 1-\left( t+t_{0}\right) \frac{%
b^{\prime}}{b}\right) y\partial_{y}+\left( 1-\left( t+t_{0}\right) \frac{%
b^{\prime}}{b}\right) z\partial_{z},
\end{equation}
where the scale factors behave as%
\begin{equation}
b(t)=b_{0}\left( t+t_{0}\right) \qquad\text{and}\qquad a(t)=a_{0}\left(
t+t_{0}\right) ^{m},\qquad\forall m\in\mathbb{R}^{+}.  \label{restric}
\end{equation}

Tanking into account these result we shall calculate some exact cosmological
solutions. It is a straightforward task to carry out the calculations for
this reason we only show the results. The detailed exposition of the
followed method may be found for example in \cite{Tony1}.

\subsection{Vacuum solution}

There is no SS vacuum solution for this model

\subsection{Perfect fluid model}

The stress-energy tensor is defined by, $T_{ij}=\left( \rho+p\right)
u_{i}u_{j}-pg_{ij},$ and where we are taking into account the conservation
principle so, $T_{ij}^{;j}=0,$ and the equation of state $%
p=\gamma\rho,\left( \gamma=const.\right) $. We find that

\begin{equation}
a(t)=a_{0}\left( t+t_{0}\right) ^{\sqrt{2}},\qquad b(t)=b_{0}\left(
t+t_{0}\right) ,\qquad \rho =\rho _{0}\left( t+t_{0}\right) ^{-2},
\end{equation}%
and this solution is only valid for
\begin{equation}
\gamma _{c}=1-\sqrt{2}\thickapprox -0.41421356.  \label{gc}
\end{equation}

Note that our solution accelerates since
\begin{equation}
q=\frac{3}{h}-1=\frac{3}{2+\sqrt{2}}-1<0.
\end{equation}

\subsection{Time-varying constant model}

In this model we consider the constants $G$ and $\Lambda$ as time varying
functions in such a way that the FE are:%
\begin{equation}
G_{ij}=G(t)T_{ij}-\Lambda(t)g_{ij},\qquad\left( G(t)T_{ij}-\Lambda
(t)g_{ij}\right) ^{;j}=0,
\end{equation}
with the constrain $T_{ij}^{;j}=0.$

We have found the next results%
\begin{align}
a(t)& =a_{0}\left( t+t_{0}\right) ^{\sqrt{2}},\qquad b(t)=b_{0}\left(
t+t_{0}\right) ,\qquad \rho =\rho _{0}\left( t+t_{0}\right) ^{-\left( \gamma
+1\right) \left( 2+\sqrt{2}\right) }  \notag \\
G& =G_{0}\left( t+t_{0}\right) ^{\left( \gamma +1\right) \left( 2+\sqrt{2}%
\right) -2},\qquad \Lambda =\Lambda _{0}\left( t+t_{0}\right) ^{-2},
\end{align}%
where%
\begin{equation}
G\thickapprox \left\{
\begin{array}{l}
\text{decreasing }\forall \gamma \in \lbrack -1,\gamma _{c}) \\
\text{constant if }\gamma =\gamma _{c} \\
\text{increasing }\forall \gamma \in (\gamma _{c},1]%
\end{array}%
\right. \qquad \Lambda _{0}\thickapprox \left\{
\begin{array}{l}
\text{negative }\forall \gamma \in \lbrack -1,\gamma _{c}) \\
\text{vanish if }\gamma =\gamma _{c} \\
\text{positive }\forall \gamma \in (\gamma _{c},1]%
\end{array}%
\right.
\end{equation}%
where $\gamma _{c}$ is given by Eq. (\ref{gc}). This solution is valid for
all EoS $\gamma ,$ i.e. there is no restrictions on the $\gamma $ parameter.
As above $q<0.$

\subsection{JBD model with CC}

For this model (this corresponds to the exposed one in section 2) the FE are
as follows%
\begin{align}
R_{ij}-\frac{1}{2}g_{ij}R& =\frac{8\pi }{c^{4}\phi }T_{ij}+\Lambda \left(
\phi \right) g_{ij}+\frac{\omega }{\phi ^{2}}\left( \phi _{,i}\phi _{,j}-%
\frac{1}{2}g_{ij}\phi _{,l}\phi ^{,l}\right) +\frac{1}{\phi }\left( \phi
_{;ij}-g_{ij}\square \phi \right) ,  \label{cc1} \\
\left( 3+2\omega \left( \phi \right) \right) \square \phi & =8\pi T-\omega
_{\phi }\phi _{,l}\phi ^{,l}-2\phi \left( \phi \Lambda _{\phi }-\Lambda
\left( \phi \right) \right) ,
\end{align}%
where $T=T_{i}^{i}$ is the trace of the stress-energy tensor. The
gravitational coupling $G_{\mathrm{eff}}(t)$ is given by

\begin{equation}
G_{\mathrm{eff}}(t)=\left( \frac{2\omega+4}{2\omega+3}\right) \frac{G_{\ast }%
}{\phi(t)}.  \label{Gcc}
\end{equation}

From the stated theorem of section 2 we already know that the physical
quantities behave as follows:%
\begin{align*}
\phi & =\phi _{0}\left( t+t_{0}\right) ^{n},\qquad \Longrightarrow \qquad
\Lambda \left( \phi \right) =\Lambda _{0}\phi ^{\frac{-1}{n}\left( n+\alpha
\right) }=\Lambda _{0}\left( t+t_{0}\right) ^{-\left( n+\alpha \right)
},\qquad \text{\ with\qquad }n+\alpha =2. \\
\omega \left( \phi \right) & =const.,\qquad \text{\ and\qquad }\rho =\rho
_{0}\left( t+t_{0}\right) ^{-\alpha },\qquad \alpha =\left( 1+\gamma \right)
h,\qquad h=2+m.
\end{align*}

We have found the next solution. The scale factors behave as%
\begin{equation}
a(t)=a_{0}\left( t+t_{0}\right) ^{m},\qquad b(t)=b_{0}\left( t+t_{0}\right) ,
\end{equation}%
with%
\begin{equation}
m=\frac{1}{2\gamma }\left( 1-\gamma -A\right) \in \left[ 1.236\,1,4\right]
,\qquad \forall \gamma \in \left[ -1,\frac{1}{9}\right] ,\qquad m_{\gamma
=1}=0,
\end{equation}%
with $A=\sqrt{9\gamma ^{2}-10\gamma +1}.$ $m$ is not defined $\forall \gamma
\in \left( \frac{1}{9},1\right) .$
\begin{equation}
\phi =\phi _{0}\left( t+t_{0}\right) ^{n},\qquad G_{\mathrm{eff}%
}\thickapprox \phi ^{-1},
\end{equation}%
where%
\begin{align}
n& =\frac{1}{2\gamma }\left( -1-3\gamma ^{2}+A\left( 1+\gamma \right)
\right) \in \left[ -\frac{14}{3},2\right] ,\qquad \forall \gamma \in \left[
-1,\frac{1}{9}\right] ,  \notag \\
n& >0,\,\,\forall \gamma \in \left[ -1,\gamma _{c}\right) ,\qquad n_{\gamma
_{c}}=0,\qquad n<0,\,\,\forall \gamma \in \left( \gamma _{c},\frac{1}{9}%
\right] ,\qquad n_{\gamma =1}=-2,
\end{align}%
note that $\gamma _{c}$ is given by Eq. (\ref{gc}).%
\begin{align}
\phi _{0}& =-\frac{1+4\gamma ^{2}-7\gamma +A\left( 2\gamma -1\right) }{%
\gamma \left( -1+3\gamma +A\right) }\in \left[ -3,\frac{11}{3}\right]
,\qquad \forall \gamma \in \left[ -1,\frac{1}{9}\right] ,  \notag \\
\phi _{0}& <0\,\,\forall \gamma \in \left[ -1,\gamma _{1}\right) ,\qquad
\phi _{0\gamma _{1}}=0,\qquad \phi _{0}>0,\,\,\forall \gamma \in \left(
\gamma _{1},\frac{1}{9}\right] ,\qquad \phi _{0\gamma =1}=1,
\end{align}%
where $\gamma _{1}=-0.1708203932.$ Therefore this solution is only valid if $%
\gamma \in \left( \gamma _{1},\frac{1}{9}\right] $ and $\gamma =1.$
\begin{equation*}
\Lambda \left( \phi \right) =\Lambda _{0}\left( t+t_{0}\right) ^{-2},\qquad
\Lambda _{0}=\Lambda _{0}\left( \gamma ,\omega \right) ,\,\qquad \omega
\left( \phi \right) =const=10^{4},
\end{equation*}%
(the recent value of $\omega \left( \phi \right) $ has been obtained from
\cite{wv}) where the performed numerical analysis shows us that: $\Lambda
_{0}>0\,\,\forall \gamma \in \left( -1,\gamma _{2}\right) ,$ $\Lambda
_{0\gamma _{2}}=0,$ $\Lambda _{0}>0\,\,\forall \gamma \in \left( \gamma
_{2},\gamma _{3}\right) ,$ $\Lambda _{0\gamma _{c}}=0,$ $\Lambda
_{0}>0\,\,\forall \gamma \in \left( \gamma _{3},\frac{1}{9}\right] ,$ and $%
\Lambda _{0\gamma =1}=1,$ where $\gamma _{2}=-0.4142736105,$ and $\gamma
_{c}=-0.4142135624.$ Note that $\gamma _{1}\in \left( \gamma _{3},\frac{1}{9}%
\right] ,$ and then $\Lambda _{0\gamma _{1}}>0.$
\begin{equation}
\rho =\rho _{0}\left( t+t_{0}\right) ^{-\alpha },\qquad \alpha =\left(
1+\gamma \right) \left( m+2\right) ,\qquad \rho _{0}=\rho _{0}\left( \gamma
,\omega \right) ,
\end{equation}%
\begin{align}
\rho _{0}& >0\,\,\forall \gamma \in \left( -1,\gamma _{4}\right) ,\qquad
\rho _{0\gamma _{4}}=0,\qquad \rho _{0}<0\,\,\forall \gamma \in \left(
\gamma _{4},\gamma _{5}\right) ,\qquad \rho _{0\gamma _{5}}=0,  \notag \\
\rho _{0}& >0\,\,\forall \gamma \in \left( \gamma _{5},\gamma _{1}\right)
,\qquad \rho _{0\gamma _{1}}=0,\qquad \rho _{0}<0,\,\,\forall \gamma \in
\left( \gamma _{1},\frac{1}{9}\right] ,\qquad \rho _{0\gamma
_{=1}}=-795.8940815,
\end{align}%
where $\gamma _{4}=-0.4187505363,\gamma _{5}=-0.4097441700,$ and $\gamma
_{1}=-0.1708203932.$

Therefore this solution is unphysical, since $\phi _{0}>0,\,\,\forall \gamma
\in \left( \gamma _{1},\frac{1}{9}\right] $ but $\rho _{0}<0,\,\,\forall
\gamma \in \left( \gamma _{1},\frac{1}{9}\right] .$ Notice that $m_{\gamma
_{c}}=\sqrt{2},$ $n_{\gamma _{c}}=0,$ and $\Lambda _{0\gamma _{c}}=0$ as in
the above solution.

\subsection{JBD model with potential}

For this model (which corresponds to the particular model, $m=1,$ exposed in
section 3) the FE are as follows%
\begin{align}
G_{ij}& =\frac{8\pi }{c^{4}\phi }T_{ij}+\frac{\omega }{\phi ^{2}}\left( \phi
_{,i}\phi _{,j}-\frac{1}{2}g_{ij}\phi _{,l}\phi ^{,l}\right) +\frac{1}{\phi }%
\left( \phi _{;ij}-g_{ij}\square \phi \right) +\frac{U\left( \phi \right) }{%
\phi }g_{ij}, \\
\left( 3+2\omega \left( \phi \right) \right) \square \phi & =8\pi T-\omega
_{\phi }\phi _{,l}\phi ^{,l}+\phi U_{\phi }-2U\left( \phi \right) ,
\end{align}%
where $T=T_{i}^{i}$ is the trace of the stress-energy tensor. The
gravitational coupling $G_{\mathrm{eff}}(t)$ is given by

\begin{equation}
G_{\mathrm{eff}}(t)=\left( \frac{2\omega+4}{2\omega+3}\right) \frac{G_{\ast }%
}{\phi(t)}.  \label{GU}
\end{equation}

\begin{align*}
\phi & =\phi_{0}\left( t+t_{0}\right) ^{n}\qquad\Longrightarrow\qquad
U\left( \phi\right) =U_{0}\phi^{\frac{1}{n}\left( n-2\right) }=U_{0}\left(
t+t_{0}\right) ^{\left( n-2\right) },\qquad\text{\ with\qquad }%
n+\alpha=2.\qquad G_{\mathrm{eff}}\thickapprox\phi^{-1}, \\
\omega\left( \phi\right) & =const.\qquad\text{\ and\qquad}\rho=\rho
_{0}\left( t+t_{0}\right) ^{-\alpha},\qquad\alpha=\left( 1+\gamma\right)
h,\qquad h=2+m
\end{align*}

We have found the following solution. The scale factors behave as%
\begin{equation}
a(t)=a_{0}\left( t+t_{0}\right) ^{m},\qquad b(t)=b_{0}\left( t+t_{0}\right) ,
\end{equation}%
with%
\begin{equation}
m=\frac{1}{2\gamma }\left( 1-\gamma -A\right) \in \left[ 1.236\,1,4\right]
,\qquad \forall \gamma \in \left[ -1,\frac{1}{9}\right] ,\qquad m_{\gamma
=1}=0
\end{equation}%
with $A=\sqrt{9\gamma ^{2}-10\gamma +1}.$ $m$ is not defined $\forall \gamma
\in \left( \frac{1}{9},1\right) .$
\begin{equation}
\phi =\phi _{0}\left( t+t_{0}\right) ^{n},\qquad G_{\mathrm{eff}%
}\thickapprox \phi ^{-1}=G_{\ast }\left( t+t_{0}\right) ^{-n},
\end{equation}%
where%
\begin{align}
n& =\frac{1}{2\gamma }\left( -1-3\gamma ^{2}+A\left( 1+\gamma \right)
\right) \in \left[ -\frac{14}{3},2\right] ,\qquad \forall \gamma \in \left[
-1,\frac{1}{9}\right] ,  \notag \\
n& >0,\,\,\forall \gamma \in \left[ -1,\gamma _{c}\right) ,\qquad n_{\gamma
_{c}}=0,\qquad n<0,\,\,\forall \gamma \in \left( \gamma _{c},\frac{1}{9}%
\right] ,\qquad n_{\gamma =1}=-2,
\end{align}%
note that $\gamma _{c}$ is given by Eq. (\ref{gc}), while%
\begin{equation}
\phi _{0}=\frac{4\gamma \left( 3\gamma -1+A\right) }{5\gamma ^{2}+4\gamma
-1+A\left( \gamma +1\right) },\qquad \phi _{0}>0,\,\,\forall \gamma \in %
\left[ -1,\frac{1}{9}\right] ,\qquad \phi _{0\gamma =1}=1,
\end{equation}%
Therefore this solution is only valid if $\gamma \in \left[ -1,\frac{1}{9}%
\right] $ and $\gamma =1.$
\begin{equation}
U\left( \phi \right) =U_{0}\left( t+t_{0}\right) ^{n-2},\qquad
U_{0}=U_{0}\left( \gamma ,\omega \right) ,\,\qquad \omega \left( \phi
\right) =const=10^{4},
\end{equation}%
\begin{equation*}
U_{0}>0\,\,\forall \gamma \in \left( -1,\frac{1}{9}\right] \backslash \left(
\gamma _{2},\gamma _{c}\right) ,\qquad U_{0\gamma =1}=1,
\end{equation*}%
\begin{equation*}
U_{0\gamma _{2}}=0,\qquad U_{0}<0\,\,\forall \gamma \in \left( \gamma
_{2},\gamma _{c}\right) ,\qquad U_{0\gamma _{3}}=0,
\end{equation*}%
where $\gamma _{2}=-0.4142736105,$ and $\gamma _{c}=-0.4142135624.$ Note
that $\left( n-2\right) <0.$
\begin{equation}
\rho =\rho _{0}\left( t+t_{0}\right) ^{-\alpha },\qquad \alpha =\left(
1+\gamma \right) \left( m+2\right) ,\qquad \rho _{0}=\rho _{0}\left( \gamma
,\omega \right) ,
\end{equation}%
\begin{equation*}
\rho _{0}<0\,\,\forall \gamma \in \left( -1,\frac{1}{9}\right] \backslash
\left( \gamma _{4},\gamma _{5}\right) ,\qquad \rho _{0\gamma
_{=1}}=-795.8940815,
\end{equation*}%
\begin{equation*}
\rho _{0\gamma _{4}}=0,\qquad \rho _{0}>0\,\,\forall \gamma \in \left(
\gamma _{4},\gamma _{5}\right) ,\qquad \rho _{0\gamma _{5}}=0,
\end{equation*}%
where $\gamma _{4}=-0.4187505363,\gamma _{5}=-0.4097441700.$

\begin{figure}[h!]
\begin{center}
\includegraphics[
height=1.9993in, width=3.0007in
]{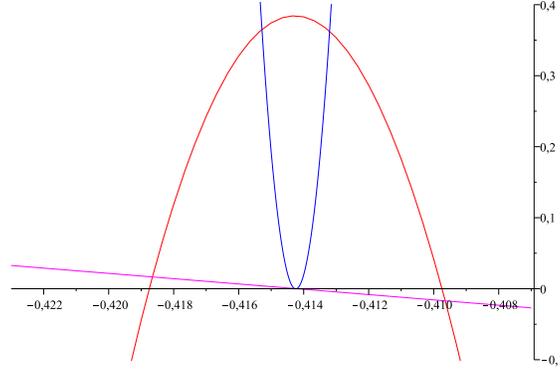}
\end{center}
\caption{JBD model with potential. Solution $\forall \protect\gamma \in
\left( \protect\gamma _{4},\protect\gamma _{5}\right)$. $\protect\rho_{0}$
is plotted in red color. $U_{0}$ in blue and $n$ in magenta color.}
\label{JBDU}
\end{figure}

Therefore this solution is only valid $\forall \gamma \in \left( \gamma
_{4},\gamma _{5}\right) \ni \gamma _{c},$ where $\rho _{0}>0$ and $\phi
_{0}>0$ while $U_{0}>0\,\,\forall \gamma \in \left( \gamma _{4},\gamma
_{5}\right) \backslash \left( \gamma _{2},\gamma _{c}\right) $, see fig. (%
\ref{JBDU})$.$ Notice that $G_{\mathrm{eff}}\left( \gamma \right) $ is
decreasing if $\gamma \in \left( \gamma _{4},\gamma _{c}\right) ,$ constant
if $\gamma =\gamma _{c}$ and growing if $\gamma \in \left( \gamma
_{c},\gamma _{5}\right) .$ As in the above solutions we have that that $%
m_{\gamma _{c}}=\sqrt{2},$ $n_{\gamma _{c}}=0,$ and $\Lambda _{0\gamma
_{c}}=0.$ We also emphasize that $q<0,$ $\forall \gamma \in \left( \gamma
_{4},\gamma _{5}\right) .$ To end we have calculated the values of $m$ in $%
\gamma _{4}$ and $\gamma _{5},$ they are: $m_{\gamma
_{4}}=1.4117477876367817948$ and $m_{\gamma _{5}}=1.4166732719815729598.$
These values will acquire a complete sense in the next model.

\subsection{Chameleon JBD\ model}

This model corresponds to the exposed one in section 4. In this case FE read:%
{\
\begin{align}
G_{ij}& =\frac{J(\phi )}{\phi }T_{ij}+\frac{\omega }{\phi ^{2}}\left( \phi
_{,i}\phi _{,j}-\frac{1}{2}g_{ij}\phi _{,l}\phi ^{,l}\right) +\frac{1}{\phi }%
\left( \phi _{;ij}-g_{ij}\square \phi \right) +\frac{U\left( \phi \right) }{%
\phi }g_{ij}, \\
\left( 2\omega +3\right) \Box \phi & =T\left( J-\frac{1}{2}\phi J_{\phi
}\right) -(\phi U_{\phi }-2U),
\end{align}%
}where the main quantities behave as follows%
\begin{align}
\phi & =\phi _{0}\left( t+t_{0}\right) ^{n},\qquad U\left( \phi \right)
=U_{0}\phi ^{\frac{1}{n}\left( n-2\right) }=U_{0}\left( t+t_{0}\right)
^{\left( n-2\right) },\qquad G_{\mathrm{eff}}\thickapprox \phi ^{-1},  \notag
\\
\omega \left( \phi \right) & =const.\qquad \text{\ and\qquad }\rho =\rho
_{0}\left( t+t_{0}\right) ^{-\alpha },\qquad \alpha =\left( 1+\gamma \right)
h,\qquad h=2+m, \\
J\left( \phi \right) & =J_{0}\phi ^{\frac{\alpha +n-2}{n}}=J_{0}\left(
t+t_{0}\right) ^{\alpha +n-2}.  \notag
\end{align}

We have obtained the following solution. $n=n(m);$%
\begin{equation}
n=-\frac{m^{2}-2}{m-1},
\end{equation}%
in such a way that $n=0,$ iff $m_{c}=\pm 1.414213562=\pm \sqrt{2},$ we only
consider the positive solution, so, $m\in \mathbb{R}^{+}\backslash \left\{
1\right\} .$ $n<0$ if $m\in \left[ 0,1\right) \cup \left( m_{c+},\infty
\right) ,$ and $n>0$ if $m\in \left( 1,m_{c+}\right) .$

The rest of the quantities depend on $\left( m,\gamma ,\omega \right) ,$ so
in order to carry out the numerical analysis it is necessary to fix the
value of $\omega $ and $\gamma .$ Setting $\omega =4\cdot 10^{4},$ we have
studied some equation of state, $\gamma =1,1/3,0,-1/3,$ i.e. the usual ones.

For $\gamma =1$ then; $\phi =\phi _{0}\left( t+t_{0}\right) ^{n},$ and then,
$G_{\mathrm{eff}}\thickapprox \phi ^{-1}=G_{\ast }\left( t+t_{0}\right)
^{-n},$ where
\begin{equation}
\phi _{0}=\phi _{0}\left( m,\gamma =1,\omega \thickapprox 10^{4}\right) =-%
\frac{(10002m^{4}-2m^{3}-60007m^{2}+80010)}{(5001m^{4}-m^{3}-9999m^{2}-3m+2)}%
,
\end{equation}%
then%
\begin{equation}
\phi _{0}=0,\qquad \Longleftrightarrow \qquad m_{1}=1.414184279,\quad
m_{2}=2.000024995,
\end{equation}%
and $\phi _{0}$ is not defined when%
\begin{equation*}
m_{a_{1}}=0.1399429180,\quad m_{a_{2}}=1.414180724,
\end{equation*}%
in such a way that $\phi _{0}>0$ if $m\in \left( m_{a_{1}},m_{a_{2}}\right)
. $ $\phi _{0}<0$ if $m\in \left( m_{a_{2}},m_{1}\right) ,$ and $\phi _{0}>0$
if $m\in \left( m_{1},m_{2}\right) .$ $U_{0}$ behaves as%
\begin{equation}
U_{0}=\phi _{0}\frac{\left( m^{2}-2m+2\right) }{2\left( m-1\right) ^{2}},
\end{equation}%
therefore $U_{0}$ has the same roots than $\phi _{0}$ and it is not defined
when%
\begin{equation*}
m_{a_{1}}=0.1399429180,\quad m_{a_{3}}=1,\quad m_{a_{2}}=1.414180724,
\end{equation*}%
in such a way that $U_{0}>0$ if $m\in \left( m_{a_{1}},m_{a_{2}}\right)
\backslash \left\{ m_{a_{3}}\right\} .$ $U_{0}<0$ if $m\in \left(
m_{a_{2}},m_{1}\right) ,$ and $U_{0}>0$ if $m\in \left( m_{1},m_{2}\right) .$

$\rho _{0}=\rho _{0}\left( m,\gamma =1/3,\omega \thickapprox
10^{4},J_{0}=10\right) $ behaves as%
\begin{equation*}
\rho _{0}=\phi _{0}\frac{(10002m^{4}-2m^{3}-40003m^{2}-2m+40006)}{160\pi
\left( 5001m^{4}-m^{3}-9999m^{2}-3m+2\right) \left( m-1\right) ^{2}},
\end{equation*}%
then $\rho _{0}=0,$ iff%
\begin{equation*}
m_{1}=1.414184279,\quad m_{2}=2.000024995,\quad m_{3}=1.411747788,\quad
m_{4}=1.416673272,
\end{equation*}%
and it is not defined if%
\begin{equation*}
m_{a_{1}}=0.1399429180,\quad m_{a_{3}}=1,\quad m_{a_{2}}=1.414180724,
\end{equation*}%
finding therefore that $\rho _{0}<0,$ if $m\in \left(
m_{a_{1}},m_{a_{2}}\right) \backslash \left\{ m_{a_{3}}\right\} ,$ $\rho
_{0}>0,$ $\forall m\in \left( m_{3},m_{4}\right) $ and $\rho _{0}<0,$ if $%
m\in \left( m_{4},m_{2}\right) .$ Nevertheless a careful analysis shows us
that $\rho _{0}$ is not defined when $m_{a_{2}}=1.414180724,$ note that $%
m_{a_{2}}\in \left( m_{3},m_{4}\right) .$ Thus $\rho _{0}>0,$ $\forall m\in
\left( m_{3},m_{a_{2}}\right) \cup \left( m_{1},m_{4}\right) ,$ if $m\in
\left( m_{a_{2}},m_{1}\right) $ then $\rho _{0}<0.$

Note that%
\begin{align*}
m_{a_{1}} &
=0.1399429180<m_{a_{3}}=1<m_{3}=1.411747788<m_{a_{2}}=1.414180724< \\
&
<m_{1}=1.414184279<m_{c_{+}}=1.414213562<m_{4}=1.416673272<m_{2}=2.000024995.
\end{align*}

\begin{figure}[h]
\begin{center}
\includegraphics[
height=1.9993in,
width=3.0007in
]{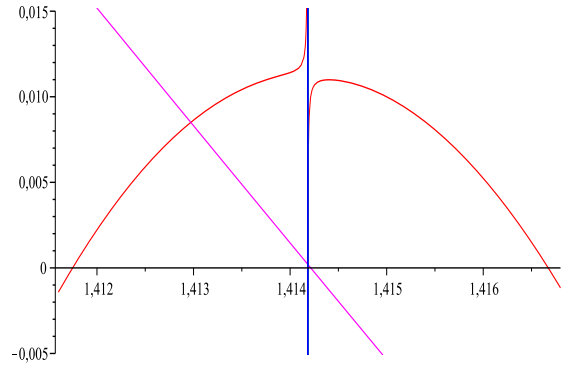}
\end{center}
\caption{Chameleon JBD model with $\protect\gamma=1$ and $m\in \left(
m_{3},,m_{4}\right)$. $\protect\rho _{0}$ is plotted in red color. $U_{0}$
in blue and $n$ in magenta color. $\protect\phi _{0}$ is plotted in green
color but it appears under the graph of $U_{0}.$}
\label{JBDC0}
\end{figure}

Therefore this solution is only valid if $m\in \left( m_{3},m_{a_{2}}\right)
\cup \left( m_{1},m_{4}\right) ,$ since in this interval $\rho _{0}>0,\phi
_{0}>0.$ In fig. (\ref{JBDC0}) we have plotted this situation.

In order to clarify and to compare (with the following results) the results
we have write in the following table the roots of the constants $\left(
n,\phi _{0},U_{0},\rho _{0}\right) :$

\begin{equation*}
\begin{array}{c||c|c|c|c|}
& r_{1} & r_{2} & r_{3} & r_{4} \\ \hline\hline
n &  &  & m_{c}=1.414213562 &  \\ \hline
\phi_{0} &  & m_{1}=1.414184279, &  &  \\ \hline
U_{0} &  & m_{1}=1.414184279, &  &  \\ \hline
\rho_{0} & m_{3}=1.411747788 & m_{1}=1.414184279, &  & m_{4}=1.416673272 \\
\hline
\end{array}%
\end{equation*}

\begin{figure}[h!]
\begin{center}
\includegraphics[
height=1.9993in,
width=3.0007in
]{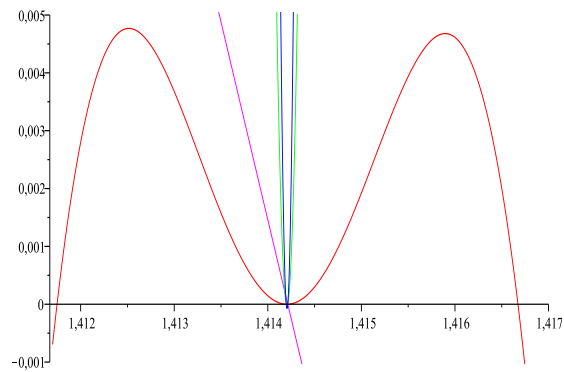}
\end{center}
\caption{Chameleon JBD model with $\protect\gamma=1/3$ and $m\in \left(
m_{3},,m_{4}\right)$. $\protect\rho_{0}$ is plotted in red color. $U_{0}$ in
blue and $n$ in magenta color. $\protect\phi_{0}$ is plotted in green color.}
\label{JBDC2}
\end{figure}

The cases $\gamma =1/3,0$ and $-1/3$, are quite similar. For example if $%
\gamma =1/3,$ the performed numerical analysis shows us that the roots of
the constants are as follows (compare with the above table)
\begin{equation*}
\begin{array}{c||c|c|c|c|}
& r_{1} & r_{2} & r_{3} & r_{4} \\ \hline\hline
n &  &  & m_{c}=1.414213562 &  \\ \hline
\phi _{0} &  & m_{1}=1.414197143 & m_{c}=1.414213562 &  \\ \hline
U_{0} &  & m_{1}=1.414197143, & m_{c}=1.414213562 &  \\ \hline
\rho _{0} & m_{3}=1.411747788 & m_{1}=1.414197143, & m_{c}=1.414213562 &
m_{4}=1.416673272 \\ \hline
\end{array}%
\end{equation*}%
and therefore we have the following scenario, see fig. \ref{JBDC2} where we
have plotted the interval $m\in \left( m_{3},,m_{4}\right)$. Note the
analogies with regard to the case $\gamma=1$.

\begin{figure}[h!]
\begin{center}
\includegraphics[
height=1.9993in,
width=3.0007in
]{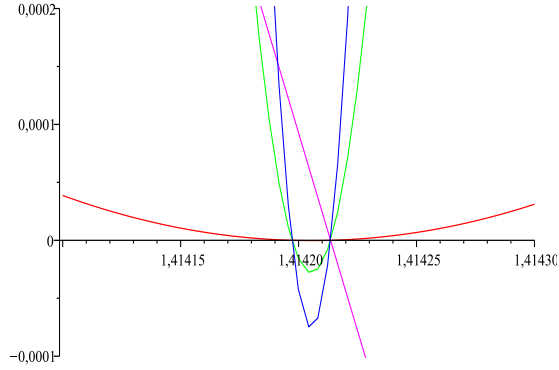}
\end{center}
\caption{Chameleon JBD model with $\protect\gamma=1/3$ and $m\in \left(
m_{1},,m_{c}\right)$. $\protect\rho_{0}$ is plotted in red color. $U_{0}$ in
blue and $n$ in magenta color. $\protect\phi_{0}$ is plotted in green color.
}
\label{JBDC5}
\end{figure}

In fig. \ref{JBDC5} we have plotted in detail the interval $m\in \left(
m_{1},,m_{c}\right)$.

Therefore this solution is valid for any value of $\gamma$, while in the
above solution it was only valid for a small interval of $\gamma_{c}$.
Remember the solution in the last model, when $m_{%
\gamma_{4}}=1.4117477876367817948$ and $m_{%
\gamma_{5}}=1.4166732719815729598. $

\subsection{Induced Gravity model}

In this model the FE read

\begin{align}
F\left( \phi\right) G_{ij} & =T_{ij}+\left( \phi_{,i}\phi_{,j}-\frac {1}{2}%
g_{ij}\phi_{,l}\phi^{,l}\right) +\left( F\left( \phi\right)
_{;ij}-g_{ij}\square F\left( \phi\right) \right) +U\left( \phi\right) g_{ij},
\\
2\square\phi & =RF_{\phi}-2U_{\phi},
\end{align}
where $R$ is the scalar curvature and $F\left( \phi\right) =\phi^{2}/4\omega$%
. The gravitational coupling $G_{\mathrm{eff}}(t)$ is given by

\begin{equation}
G_{\mathrm{eff}}(t)\thickapprox\frac{G_{\ast}}{\phi^{2}}.
\end{equation}

For the metric (\ref{ksmetric}) the scalar curvature is given by: $R=2\left(
\frac{a^{\prime\prime}}{a}+2\frac{b^{\prime}a^{\prime}}{ba}+2\frac {%
b^{\prime\prime}}{b}+\frac{1}{b^{2}}+\left( \frac{b^{\prime}}{b}\right)
^{2}\right) .$ As we already know, the physical quantities behave as
follows:
\begin{align*}
\phi & =\phi_{0}\left( t+t_{0}\right) ^{n},\qquad U\left( \phi\right)
=U_{0}\phi^{\frac{2}{n}\left( n-1\right) }=U_{0}\left( t+t_{0}\right)
^{2\left( n-1\right) },\qquad G_{\mathrm{eff}}\thickapprox\phi^{-2}, \\
\omega\left( \phi\right) & =\mathrm{const}.,\qquad\text{and\qquad}%
\rho=\rho_{0}\left( t+t_{0}\right) ^{-\alpha},\qquad\alpha=\left(
1+\gamma\right) h,\qquad h=2+m.
\end{align*}

We have found the following solution. The scale factors behave as%
\begin{equation*}
a(t)=a_{0}\left( t+t_{0}\right) ^{m},\qquad b(t)=b_{0}\left( t+t_{0}\right) ,
\end{equation*}
with $m=m(n):$%
\begin{equation*}
m=-n+A,\qquad A=\sqrt{2n^{2}+n+2},
\end{equation*}
where, as it is observed if $n=0,$ then $m=\sqrt{2}.$ $\phi_{0}=\phi
_{0}\left( n,\gamma\right) $ behaves as follows%
\begin{equation*}
\phi_{0}=\frac{2n\left[ 3n^{2}\left( 1+\gamma\right) +n\left(
1-2A+\omega\left( 1+A\right) \left( 1+\gamma\right) -\left( 2A+1\right)
\gamma\right) +3\left( \gamma+1\right) \right] }{n^{3}\left(
2+\omega-\gamma\left( 1+\omega\right) \right) +n^{2}\left( \gamma\left(
2+\omega+A\right) -\omega-1\right) -\gamma\left( A+1\right) -1},
\end{equation*}
note that $\phi_{0}=0,$ iff $\gamma=-1,$and $n\neq0.$ With regard to the
constant $U_{0}=U_{0}\left( n,\gamma,\omega\right) $ we have found the next
value

\begin{equation*}
U_{0}=\frac{\phi_{0}}{\omega\left( n-1\right) }\left( n\left( A-n\right)
\left( \omega-2\right) +\left( n^{2}+1\right) \left( \omega+1\right)
+3\right) .
\end{equation*}
The constant for the energy density, $\rho_{0}=\rho_{0}\left( n,\gamma
,\omega\right) ,$ behaves as follows:%
\begin{equation*}
\rho_{0}=\frac{\phi_{0}}{2\omega\left( n-1\right) }\left( \phi_{0}\left(
n\omega-\left( 1+n\right) +\left( n-A\right) \left( 1-n^{2}\right) \right)
-2n^{3}\left( \omega+1\right) +2n^{2}\left( -1+2\left( A-n\right) \left(
1-\omega\right) -\omega\right) -6n\right) .
\end{equation*}

Since the solutions depend on the parameters $\left( n,\gamma ,\omega
\right) $ then we need to fix them. For example, if we set For $\omega
\thickapprox 4\cdot 10^{4},$ then the solutions only depend on $\left(
n,\gamma \right) $ in such a way that given different values to $n$ then we
shall may to study their behaviour. For $n=-1,$ $\omega \thickapprox 4\cdot
10^{4}$ we get $\phi _{0}>0$ and $\rho _{0}>0$ iff $\gamma <-1$ while $%
U_{0}<0$ if $\gamma <-1.$ We arrive at the same conclusion if $n=-2.$ Note
that the cases $n=0$ and $n=1$ are forbidden. Therefore, the obtained
solution is only valid if $\gamma <-1.$

\section{Conclusions}

We have studied under the self-similar hypothesis the admitted form of the
different unknown functions in several scalar-tensor theories. By employing
the matter collineation (MC) approach, i.e. calculating the Lie derivate of
the effective stress-energy tensor with respect to an HVF, and the Lie group
method, we have been able to state theorems valid for all the Bianchi
geometries as well as for a flat FRW metric. We have used both tactics,
because with the MC in some of the models studied we are only able to obtain
relationships between the physical quantities. With this tactic we only
obtain self-similar solutions. Nevertheless, with the Lie group method we
are able to obtain the exact form for each of the physical quantities.
Furthermore, with this approach we obtain scaling solutions which are more
general than the self-similar one.

In the first of the models studied we arrive to the conclusion that $\phi
\thickapprox \left( t+t_{0}\right) ^{n},$ and therefore $G_{\mathrm{eff}%
}\thickapprox \left( t+t_{0}\right) ^{-n},$ while the dynamical cosmological
constant behaves as $\Lambda \thickapprox \left( t+t_{0}\right) ^{-2}.$ In
the same way we have deduced that the Brans-Dicke parameter $\omega
_{BD}\left( \phi \right) $ must be constant. In the second of the models
studied, the generalized scalar-tensor model, the dynamical cosmological
constant is mimicked by the potential $U\left( \phi \right) ,$ and therefore
we have three unknown functions, $F\left( \phi \right) ,$ $Z\left( \phi
\right) $ and the potential $U\left( \phi \right) .$ We arrive at a very
general result which allows us to outline different scalar-tensor models
that admit self-similar solutions. Actually this result is in agreement with
the fact that we may pass from one model to another through conformal
transformations. As an example we have emphasized three relevant models, the
standard scalar-tensor model, the induced gravity model and a specific model
which is very similar to the scalar cosmological model where $G_{\mathrm{eff}%
}\thickapprox \mathrm{const}.$ In the third of the models studied, following
the same procedure, we calculate the admitted form for the unknown functions
$J\left( \phi \right) $ and the potential in order to obtain scaling and
self-similar solutions.

Once we have established all these results then we study a particular
example, the Kantowski-Sach model. The same procedure may be applied to
other Bianchi models. We begin by showing that this metric admits and HVF.
We explore some cosmological models and applying the stated theorems we find
exact solutions. We show that there is no vacuum solution. For the perfect
fluid case, within the general relativity framework, we find that the
solution obtained is only valid for a particular equation of state, $\gamma
_{c}$, which is not strange in this class of solutions, while the exponent
of one of the scale factors is irrational. This fact is odd since these
constants usually are rational ones. Note that the solution is inflationary
since $q<0$ without any necessity to appeal to a DM component. In the third
model, where $G$ and $\Lambda $ are considered as time-varying within the
general relativity framework we have shown that the obtained solution is
valid for all value of $\gamma $. In this case $G$ may be a growing or
decreasing function finding that it behaves as a true constant only when $%
\gamma =\gamma _{c}.$ The dynamical cosmological constant is always a
decreasing time function but it may be positive or negative and it vanishes
if $\gamma =\gamma _{c}.$ Once we know how each physical quantity works in
the general relativity framework then we explore some solutions for the
scalar-tensor models.

We have not been able to find a solution in the case of a Brans-Dicke model
with a dynamical $\Lambda .$ This result is quite surprising since following
the same procedure we have obtained solutions for several Bianchi models.
Nevertheless for the BD model with a potential we obtain an exact
self-similar solution. The numerical analysis carried out shows us that the
solution obtained is only valid in a small neighbourhood, $\mathcal{E}\left(
\gamma _{c}\right) ,$ of $\gamma _{c},$ the critical value of $\gamma ,$
where the energy density and the scalar function are positive. The solution
shows analogies with that obtained in the subsection where $G$ and $\Lambda $
are considered as time-varying within the general relativity framework. For
example, $G_{\mathrm{eff}}\thickapprox \left( t+t_{0}\right) ^{-n}$ may be a
growing or decreasing function finding that it behaves as a true constant
only when $\gamma =\gamma _{c}.$ The dynamical cosmological constant, the
potential $U\left( \phi \right) ,$ is always a decreasing time function but
it may be positive in the interval of definition or vanish if $\gamma
=\gamma _{c}.$ The model also accelerates since $q<0.$ Trying to generalize
this scenario we also consider a chameleon BD model. We show that the
obtained solution is valid for all values of $\gamma $ instead of only in $%
\mathcal{E}\left( \gamma _{c}\right) .$ In the last model studied, the
induced gravity case, we find that the solution is only valid if $\gamma <-1$
(a phantom scenario). In the appendix we emphasize the fact that the only
form compatible with self-similar solutions for the BD parameter $\omega
_{BD}\left( \phi \right) $ is $\omega _{BD}\left( \phi \right) =\mathrm{const%
}.$ Therefore none of the scaling solutions admit a variable $\omega
_{BD}\left( \phi \right) .$

\appendix

\section{Study of Eq. (\protect\ref{Eq phi})}

In this appendix we shall study through the Lie group method the Eq. (\ref%
{Eq phi}) i.e.%
\begin{equation}
\phi _{tt}=-\frac{\phi _{t}^{2}}{\phi }\left( \frac{1}{2}\frac{\omega _{\phi
}}{\omega }\phi -1\right) -\frac{\phi _{t}}{t}.  \label{app1}
\end{equation}

Therefore, following the standard procedure, we need to solve the next
system of PDE:

\begin{eqnarray}
W\xi _{\phi }-\phi \xi _{\phi \phi } &=&0, \\
-W\eta +\frac{\phi }{2\omega }\left( \omega _{\phi \phi }\phi +\omega _{\phi
}\left( 1-\frac{\omega _{\phi }}{\omega }\phi \right) \right) \eta
+2t^{-1}\phi ^{2}\xi _{\phi }+W\phi \eta _{\phi }+\phi ^{2}\eta _{\phi \phi
}-2\phi ^{2}\xi _{t\phi } &=&0, \\
t^{-2}\left( t\xi _{t}-\xi \right) +2W\phi ^{-1}\eta _{t}+2\eta _{t\phi
}-\xi _{tt} &=&0, \\
t^{-1}\eta _{t}+\eta _{tt} &=&0,
\end{eqnarray}%
where $W=\frac{1}{2}\frac{\omega _{\phi }\phi }{\omega }-1.$

The symmetry $\xi =t,$ $\eta =n\phi $, brings us to obtain the following
constrain on the function $\omega \left( \phi \right) :$%
\begin{equation*}
-W\eta +\frac{\phi }{2\omega }\left( \omega _{\phi \phi }\phi +\omega _{\phi
}\left( 1-\frac{\omega _{\phi }\phi }{\omega }\right) \right) \eta +W\phi
\eta _{\phi }=0,
\end{equation*}%
i.e.%
\begin{equation*}
-\left( \frac{1}{2}\frac{\omega _{\phi }\phi }{\omega }-1\right) n\phi +%
\frac{\phi }{2\omega }\left( \omega _{\phi \phi }\phi +\omega _{\phi }\left(
1-\frac{\omega _{\phi }\phi }{\omega }\right) \right) n\phi +\left( \frac{1}{%
2}\frac{\omega _{\phi }\phi }{\omega }-1\right) \phi n=0,
\end{equation*}%
and therefore we get%
\begin{equation*}
\omega _{\phi \phi }\phi +\omega _{\phi }\left( 1-\frac{\omega _{\phi }\phi
}{\omega }\right) =0,
\end{equation*}%
so%
\begin{equation*}
\omega _{\phi \phi }=\frac{\omega _{\phi }^{2}}{\omega }-\frac{\omega _{\phi
}}{\phi }\qquad \Longleftrightarrow \qquad \omega \left( \phi \right)
=\omega _{0}\phi ^{\delta },\qquad \omega _{0},\delta \in \mathbb{R}.
\end{equation*}

Therefore if $\phi =\phi _{0}t^{n},$ then $\omega \left( \phi \right)
=\omega _{0}\phi ^{\delta }=\omega _{0}t^{n\delta }.$

If we substitute this result into Eq. (\ref{app1}) then we get%
\begin{equation*}
n\left( n-1\right) =-n^{2}\left( \frac{1}{2}\delta -1\right) -n\qquad
\Longleftrightarrow \qquad \delta =0,
\end{equation*}%
this means that
\begin{equation*}
\omega \left( \phi \right) =const.
\end{equation*}%
as we already know.

\end{document}